\documentclass[aps,preprint,showpacs,nofootinbib]{revtex4}

\usepackage[utf8]{inputenc}
\usepackage{color}
\usepackage{amsmath,amssymb}
\usepackage{amsfonts}
\usepackage{verbatim}
\usepackage{makeidx}
\usepackage{hyperref}
\usepackage{slashed}
\usepackage[vcentermath]{youngtab}
\usepackage{float}
\usepackage{booktabs}
\usepackage{graphicx}
\usepackage{mathrsfs}
\usepackage{bm}
\usepackage{braket}
\usepackage{url}
\usepackage{etoolbox}
\apptocmd{\sloppy}{\hbadness 10000\relax}{}{}
\allowdisplaybreaks[3]
\def\arraystretch{1.0}

 \newcommand{\be}{\begin{equation}}
   \newcommand{\ee}{\end{equation}}
     \newcommand{\bea}{\begin{eqnarray}}
   \newcommand{\eea}{\end{eqnarray}}

\usepackage[normalem]{ulem}  
\usepackage{color}
\renewcommand\sout{\bgroup \color{red} \ULdepth=-.5ex \ULset}

\begin{document}

\title{A study of $c c\bar{c}\bar{c}$ tetraquark decays in 4 muons and 
in  $D^{(*)} \bar{D}^{(*)}$ at  LHC}

\author{C. Becchi$^1$}
\author{J. Ferretti$^2$}
\author{A. Giachino$^1$}
\author{L. Maiani$^3$}
\author{E. Santopinto$^1$}

\affiliation{$^1$Istituto Nazionale di Fisica Nucleare (INFN), Sezione di
Genova, via Dodecaneso 33, 16146 Genova, Italy
}

\affiliation{$^2$Department of Physics, University of Jyv\"askyl\"a, P.O. Box 35, 40014 Jyv\"askyl\"a, Finland}

\affiliation{$^3$Dipartimento di Fisica and INFN,  Sapienza  Universit\`a di Roma, Piazzale Aldo Moro 2, I-00185 Roma, Italy}
\date{\today}

\begin{abstract}
We perform a quantitative analysis of the decays of $cc\bar c\bar c$ tetraquarks with $J^{PC}=0^{++}, 2^{++}$ into 4 muons and into hidden- and open-charm mesons and estimate, for the first time, the fully charmed tetraquark decay width.
The calculated cross section upper limit is $\sim 40(240)$ fb  for the 4 muons channel, and $\sim 75(110)$ pb for the $D^{(*)} \bar D^{(*)} \to e\mu$ channel, in the $0^{++}(2^{++})$ case. Decay widths depend upon the additional parameter $\xi=|\Psi_{\cal T}(0)|^2/|\Psi_{J/\psi}(0)|^2$, which can be computed with a considerable error. We find $\Gamma(0^{++})=97\pm30$~MeV and $\Gamma(2^{++})=64\pm 20$~MeV. On the basis of our results and with the present sensitivity, LHCb should detect both $0^{++}$ and $2^{++}$ fully-charmed tetraquarks.
\end{abstract} 


  \maketitle

\section{\bf{Introduction}}\label{sec1}

In this note we consider production and decay at proton colliders  of the fully charmed tetraquarks, ${\cal T}=cc\bar c\bar c$. In particular, we consider the $4\mu$ and meson-meson decays, the latter revealed through the $e\mu$ signature of their weak decays. We focus on the ground states with $J^{PC}=0^{++}, 2^{++}$. We shall use the method recently applied to production and decay of fully bottom tetraquarks, $bb \bar b\bar b$ in~\cite{Becchi:2020mjz}, briefly described in the following. 

Evidence for a $4\mu$ resonance has been announced in a recent paper of the LHCb Collaboration~\cite{Aaij:2020fnh}, which is in line with our estimates and indicates that $4\mu$ and meson-meson channels may be the key to the study of these truly exotic hadrons.

The hypothetical existence of hadronic states with more than minimal quark content ($q\bar{q}$ or $qqq$) was proposed by Gell-Mann in 1964  \cite{GellMann:1964nj} and Zweig \cite{Zweig:1964}, followed by a quantitative model  by Jaffe \cite{Jaffe:1976ig} for the lightest scalar mesons described as  diquark anti-diquark pairs.
Recent years have seen  considerable growth in the observation of  four valence quark  states that cannot be included in the well-known systematics of $ q\bar q$ mesons, like $Z(4430)$~\cite{Choi:2007wga,Aaij:2014jqa} and $Z(4248)$~\cite{Ablikim:2013wzq}. Similar particles have also been found in the bottom sector, $Z_b(10610)$ and $Z_b(10650)$, observed by the Belle collaboration \cite{Bondar}  (see~\cite{Ali:2019roi,Esposito:2016noz,Lebed:2016hpi,Olsen:2017bmm,Guo:2017jvc,Liu:2019zoy} for recent reviews).

Earlier predictions of a 
fully-charmed $c c \bar{c} \bar{c} $ tetraquark were made in Refs.~\cite{Iwasaki:1975pv,Chao:1980dv,Heller:1985cb,Barnea:2006sd,Vijande:2007ix,Ebert:2007rn,Berezhnoy:2011xn}, 
 followed by more recent
~ studies in~
\cite{Wu:2016vtq,Chen:2016jxd,Bai:2016int,Wang:2017jtz,Debastiani:2017msn,Richard:2017vry,Anwar:2017toa,Richard:2018yrm,Esposito:2018cwh,Liu:2019zuc,Bedolla:2019zwg}.

Refs. \cite{Anwar:2017toa,Karliner:2016zzc} have estimated the $J^{PC}=0^{++}$, fully-bottom tetraquark decay width. 

Theoretically, $J^{PC}=0^{++}$ is expected for the $c c \bar{c} \bar{c}$ ground-state. Following Ref.~\cite{Becchi:2020mjz} we present a calculation of decay widths and branching ratios of the main, hidden- and open-charm channels of $cc\bar c\bar c$ tetraquarks. 

 To be explicit, we assume that such states do indeed exist, as in~\cite[Table III]{Bedolla:2019zwg}. \footnote{
The spectrum given in~\cite{Bedolla:2019zwg} has to be shifted by a constant $\Delta E$, determined so as to reproduce the experimental mass of the tetraquark ground-state, see~\cite{Maiani:2020pur}.}

We restrict, for definiteness, to diquarks in color ${\bar{\bf 3}}$. The case of color 6 diquarks can be worked out as a simple extension. 

We extend to fully charm tetraquarks recent work on doubly heavy tetraquarks in the quark model~\cite{Karliner:2017qjm,Eichten:2017ffp,Maiani:2019lpu} and in Lattice QCD~\cite{Junnarkar:2018twb}. It is worth noticing that in the last reference no evidence was found of bound doubly heavy diquarks in color ${\bf 6}$. The presence of color ${\bf 6}$ diquarks  component in the doubly heavy tetraquark has been noted in~\cite{Hernandez:2019eox} and found to vanish for increasing heavy to light quark mass ratio. 

In a recent paper tetraquarks with ${\bar{\bf 3}}$ diquarks and ${\bar {\bf 6}}$ antidiquarks have also been considered, in the presence of explicit gluon fields~\cite{Giron:2020wpx}. This situation is definitely beyond reach of our method.

\section{Results}


Decay rates are proportional to the ratio of overlap probabilities of the annihilating $c \bar c$ pairs in $\cal T$ and $J/\psi$:
 \be
 \xi=\frac{|\Psi_{\cal T}(0)|^2}{ |\Psi_{J/\psi}(0)|^2} 
 \ee
 Branching ratios do not depend upon $\xi$, our predictions are reported in Table~\ref{uno}. In particular, we find 
\bea
&& B({\cal T}\to 4\mu)=2.7\cdot 10^{-6}~(J^{PC}=0^{++});\notag \\
&& B({\cal T}\to 4\mu)=16\cdot 10^{-6}~(J^{PC}=2^{++}).
\label{B4mufin} 
\eea

The total width is expressed as: 
\be
\Gamma({\cal T}(J=0^{++}))=21\cdot \xi~~{\rm MeV}\notag \\
\label{widths1}
\ee
The overlap functions  for tetraquark and $J/\psi$ can be computed by making use of a variational method with harmonic oscillator trial wave functions. 
To get the  overlap function of the $J/\psi$ one can also use the leptonic width. In Sect.~\ref{value} we find:
\be
\xi=4.6\pm1.4\label{bestratio}
\ee
Our best estimate is then
\be
\Gamma({\cal T}(J=0^{++})=97\pm 30~{\rm MeV}\label{oplusplus}\\
 \ee

We extend the calculation to the $J^{PC}=2^{++}$, fully-charmed tetraquark. $J=2$ tetraquarks are produced in $p+p$ collisions with a statistical factor of $5$ with respect to the spin $0$ state. The decay  ${\cal T} \to \eta_c+{\rm light~ hadrons}$ is suppressed but annihilations into meson pairs take place  at a greater rate. 

  \begin{table}[htb!]
\centering
    \begin{tabular}{|c|c|c|c|c|c|c|}
     \hline
{\footnotesize $[cc\bar c\bar c]$}  &{\footnotesize $\eta_c$+ any} & {\footnotesize $D_q \bar D_q$ ($m_q<m_c$)} & {\footnotesize $ D^*_q \bar D^*_q$}& {\footnotesize $J/\psi$+ any} & {\footnotesize $J/\psi+\mu^+\mu^-$} &  {\footnotesize $4\mu$} \\ \hline 
 $J^{PC}=0^{++}$& {\footnotesize $0.77$}&{\footnotesize $0.019$}&{\footnotesize $0.057$}&{\footnotesize $7.5\cdot 10^{-4}$} & {\footnotesize $4.5\cdot 10^{-5}$} & {\footnotesize $2.7\cdot 10^{-6}$} \\ \hline
  $ J^{PC}=2^{++}$& 0&0&0.333& $4.4\cdot 10^{-3}$ & $2.6\cdot 10^{-4}$ &  $1.6\cdot 10^{-5}$ \\ \hline
\end{tabular}
 \caption{ \footnotesize {Branching fractions of fully-charmed tetraquarks, assuming $S$-wave decay.}}
 \label{uno}
\end{table}  
 
We  find: 
\bea
&&\Gamma({\cal T}(J=2^{++})=14\cdot \xi~~{\rm MeV}=\notag \\
&&=64\pm 20~{\rm MeV}\label{2plusplus}
\eea

Branching fractions and upper limits to the cross sections of  final states in $pp$ collisions are summarised in Tabs.~\ref{uno} and \ref{tab:width}.

The results of Tab.~\ref{uno} combined with the recent determination by LHCb of the cross section for $2  J/\psi$ production at 13 TeV~\cite{Aaij:2016bqq}, give ecouraging upper bounds to the production of ${\cal T}\to 4\mu$  at LHC
\bea
&&\sigma(p+p\to {\cal T}(0^{++})+\dots \to 4 \mu+\dots)< 40~{\rm fb}\notag \\
&&\sigma(p+p\to {\cal T}(2^{++})+\dots \to 4 \mu+\dots)< 238~{\rm fb}
\label{upp4mu}
\eea

\section{\bf{Details of the calculation}}\label{calculation}

We give here a brief description of our method. The reader may consult Ref.~\cite{Becchi:2020mjz} for more details.
The starting point is the Fierz transformation, which  brings $c \bar c$ together~\cite{Ali:2019roi}: 
\bea
&& {\cal T}(J=0^{++})= \left| \left( c^{}_{}  c^{}_{} \right)_{\bar{3}}^{\;1}  \left( \bar{c}^{}_{} \bar{c}^{}_{} \right)_{3}^{\;1}    \right \rangle_{1}^{\;0}=-\frac{1}{2}\left(    \sqrt{\frac{1}{3}} \left| \left( c^{} \bar{ c}^{} \right)^{\;1}_{1}   \left(  c^{}  \bar{ c}^{} \right)^{\;1}_{1} \right\rangle^{\;0}_{1}   -\sqrt{\frac{2}{3}} \left| \left(  c^{} \bar{ c}^{} \right)^{\;1}_{8}  \left(  c^{}  \bar{c}^{} \right)^{\;1}_{8}   \right\rangle^{\;0}_{1}  \right) + \nonumber \\ 
&&+\frac{\sqrt{3}}{2} \left(    \sqrt{\frac{1}{3}}  \left| \left(  c^{} \bar{ c}^{} \right)^{\;0}_{1}   \left(  c^{}  \bar{ c}^{} \right)^{\;0}_{1} \right\rangle^{\;0}_{1}  -\sqrt{\frac{2}{3}} \left| \left(  c^{}  \bar{ c}^{} \right)^{\;0}_{8}  \left(  c^{}  \bar{ c}^{} \right)^{\;0}_{8}   \right\rangle^{\;0}_{1}   \right).
 \label{fierz1} 
 \eea
quark bilinears are normalised to unity, subscripts denote the dimension of colour representations, and superscripts the total spin.
~For the $J=2$ tetraquark, one finds:
\bea
&& {\cal T}(J=2^{++})= \left| \left(  c^{}_{}   c^{}_{} \right)_{\bar{3}}^{\;1}  \left( \bar{ c}^{}_{} \bar{ c}^{}_{} \right)_{3}^{\;1}    \right \rangle_{1}^{\;2} =\left(    \sqrt{\frac{1}{3}}  \left| \left(  c^{} \bar{ c}^{} \right)^{\;1}_{1}   \left(  c^{}  \bar{ c}^{} \right)^{\;1}_{1} \right\rangle^{\;2}_{1}  -\sqrt{\frac{2}{3}} \left| \left(  c^{}  \bar{ c}^{} \right)^{\;1}_{8}  \left(  c^{}  \bar{ c}^{} \right)^{\;1}_{8}   \right\rangle^{\;2}_{1}   \right).
 \label{fierz2} 
 \eea

We describe ${\cal T}$ decay rate as the incoherent sum of the annhilation rates of one charm quark, call it $c_1$, with either antiquark $\bar c_1$ or $\bar c_2$(see~\cite{Becchi:2020mjz}). The $c_1$-$\bar c_1$ annihilation rate, with $c_2$ and $\bar c_2$ spectators, depends on total color and spin of the incoming particles, which are given by \eqref{fierz1} or \eqref{fierz2}. Rates are categorised in the following  items 1 to 4. Annihilation of $c_1$ with $\bar c_2$ gives the same result and brings in a factor 2. This approximation is valid in the limit of very massive quarks (mass $>>\Lambda_{QCD}\sim 0.35$~GeV), which behave as classical particles that can be localised independently.

\begin{enumerate}
\item The colour singlet, spin $0$ pair decays into  $2$  gluons,  which are converted  into confined, light hadrons with a rate  of order $\alpha_S^2$; taking the spectator $c\bar c$ pair into account,  this decay leads to: ${\cal T} \to \eta_c+{\rm light~ hadrons}$.
\item The colour singlet, spin $1$ pair decays into $3$  gluons,  which are converted  into confined light hadrons leading to:  ${\cal T} \to J/\psi+{\rm light ~hadrons}$. The rate  is of order $\alpha_S^3$. In addition, annihilation into one photon produces the final state $J/\psi +\mu^+ \mu^-$ and, eventually,  $4\mu$, with rates of order $\alpha^2$ and $\alpha^4$.
\item  The colour octet, spin $1$ pairs annihilate into one gluon, which materialises into a pair of light quark flavours, $q=u,d,s$; the latter recombine with the spectator pair to produce a pair of open-charm mesons $D_q\bar D_q$ and $D^*_q\bar D^*_q$, with a rate  of  order $\alpha_S^2$. 
\item  The colour octet, spin $0$  pairs annihilate into two gluons, which have to produce a pair of light quarks to neutralise the colour of the spectator $c\bar c$ pair, with amplitude of order  $\alpha_S^2$ and  rate  of the order of $\alpha_S^4$, which we neglect. 

\end{enumerate}

 \begin{figure}[htbp]
   \centering
         \label{graphs}
\end{figure}

The total $\cal T$ decay rate is the sum of individual decay rates, obtained from the simple formula~\cite{landlif}
\be
\Gamma((c\bar c)_c^s)=|\Psi(0)_{\cal T}|^2 v\sigma ((c\bar c)_c^s \to f)
\label{master1}
\ee
$|\Psi(0)_{\cal T}|^2$ is the overlap probability of the annihilating pair, $v$ the relative velocity, $\sigma$ the spin-averaged annihilation cross section in the final state $f$ and suffixes $s$ and $c$ denote spin and color \footnote{Our method of calculation is borrowed from the theory of $K$ electron capture, where an atomic electron reacts with a proton in the nucleus to give a final nucleus and a neutrino, see~\cite{Becchi:2020mjz}.}. For tetraquarks near the $2J/\psi$ threshold, the spectator $c \bar c$ pair appears as $\eta_c$ or $J/\psi$ on the mass shell, or  combines with the outgoing $q\bar q$ pair into an open-charm meson pair. 

We normalise the overlap probabilities to $|\Psi_{J/\psi}(0)|^2$, derived from the $J/\psi$ decay rate into lepton pairs.
Eq.~\eqref{master1} applied to this case gives:
\be
\Gamma(J/\psi \to \mu^+ \mu^-)=Q_c^2 ~\frac{4\pi \alpha^2}{3}  \frac{4}{m^2_{J/\psi}}~ |\Psi_{J/\psi}(0)|^2. \label{QMpict}
\ee
 In terms of the Vector Meson Dominance parameter~\cite{Schildknecht:2005xr} defined by
\be
J^\mu(x) =\bar c(x) \gamma^\mu c(x)=\frac{m_{J/\psi}^2}{f}~\psi^\mu (x)
\ee
with $f$ a pure number, one obtains~\cite{pdg}:
\bea
&&|\Psi_{J/\psi}(0)|^2=\frac{m_{J/\psi}^{3}}{4f^2}; \label{fandpsi} \;\; f=7.4;~|\Psi_{J/\psi}(0)|^2\sim 0.13~{\rm GeV}^{3}.\label{fandpsi2}
\eea

\emph{{\bf Numerical results.}} 
The contribution to the ${\cal T}$ decay rate of the colour singlet, spin $0$ decay  is
\bea
&&\Gamma_0=\Gamma( {\cal T}\to \eta_c + {\rm light ~hadrons})= 2\cdot \frac{1}{4} \cdot  |\Psi(0)_{\cal T}|^2 v\sigma ((c\bar c)_1^0\to 2 ~{\rm gluons})\notag \\
&&= \frac{1}{2}~\Gamma(\eta_c)\cdot \xi= 16~{\rm MeV}\cdot \xi
\eea
We have used the spectroscopic coefficient  in \eqref{fierz1} and have set
\be
|\Psi_{J/\psi}(0)|^2 v\sigma ((c\bar c)_1^0\to 2 ~{\rm gluons})\sim \Gamma(\eta_c)=32~{\rm MeV}.
\ee
Similarly
\bea
&&\Gamma_1=\Gamma( {\cal T}\to J/\psi + {\rm light ~hadrons})= 2\cdot \frac{1}{12} \cdot  |\Psi(0)_{\cal T}|^2 v\sigma ((c\bar c)_1^1\to 3 ~{\rm gluons})=\notag \\
&&= \frac{1}{6}~\Gamma(J/\psi)\cdot \xi= 16\cdot \xi~{\rm keV} \notag \\
&&\Gamma_2=\Gamma( {\cal T}\to J/\psi + \mu^+\mu^-)= B_{\mu\mu} \Gamma_1= 0.92\cdot \xi~{\rm keV}\notag \\
&& \Gamma_4=\Gamma( {\cal T}\to 4\mu)= B_{\mu\mu} ^2\Gamma_1= 5.6~10^{-2}\cdot \xi~{\rm keV} \label{gam4mu}
\eea
where $B_{\mu\mu}= B(J/\psi \to \mu^+\mu^-)$. For tetraquark mass close and above the $2J/\psi$ mass, the system recoiling against the $J/\psi$ will be dominated by a single $J/\psi$.

 \begin{figure}[htbp]
   \centering
   \includegraphics[width=0.6 \linewidth]{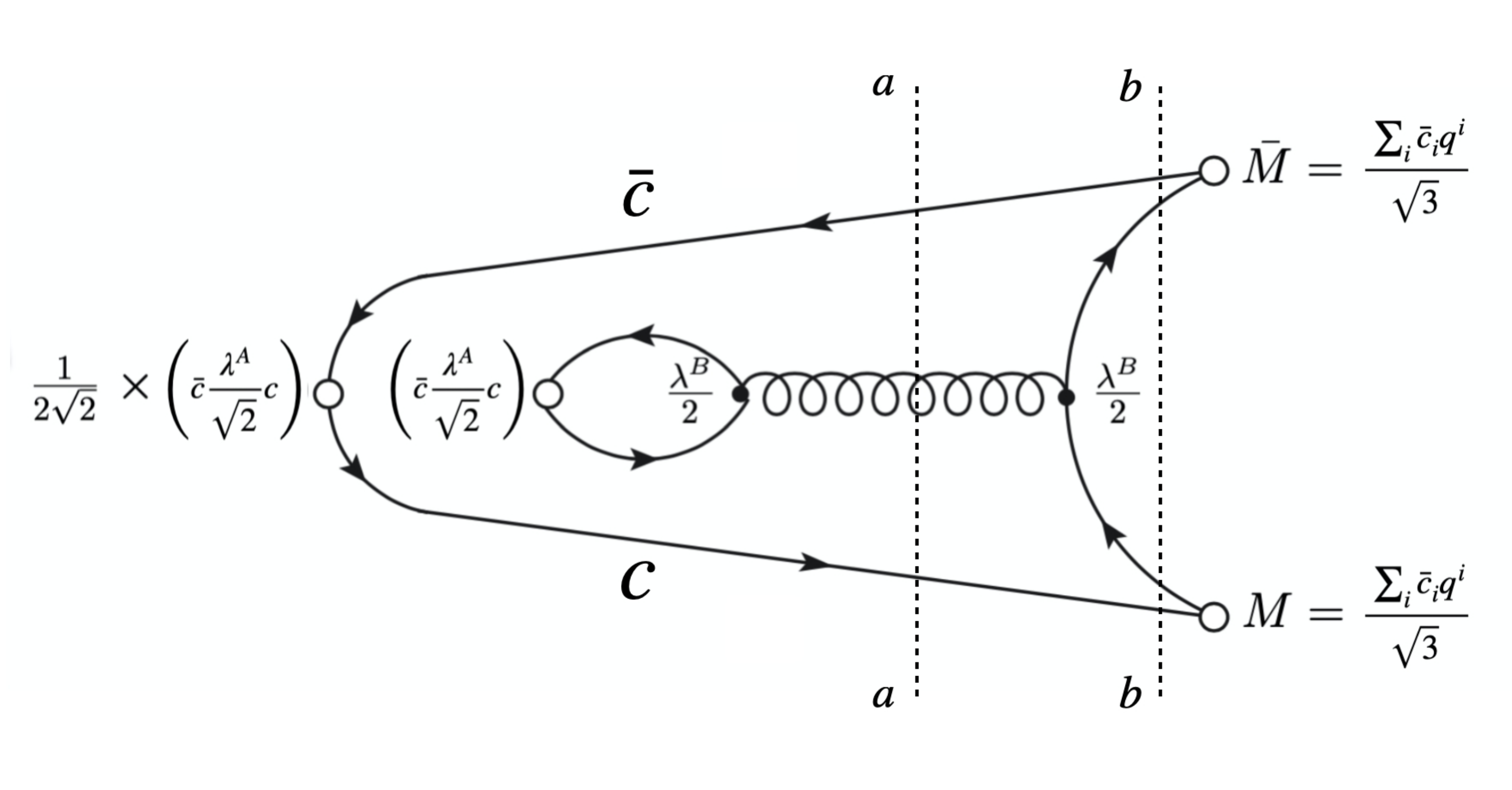}
   \caption{\footnotesize{Colour flow in $c\bar c$ annihilation. Open circles represent the insertion of quark bilinears, and black dots QCD vertices. Colour matrices and normalizations are indicated. Dotted vertical lines, $a-a$ and $b-b$, represent cuts corresponding to tetraquark decays at parton level discussed in the text.}}
         \label{chanpaton}
\end{figure}
Finally we consider the annihilation of $(c \bar c)^1_8$ into light quark pairs,  Fig.~\ref{chanpaton}. The numerical factor associated to the traces of the colour matrices along fermion closed paths, C (the Chan-Paton factor \cite{Paton:1969je}) gives the effective coupling constant of the process,
$
\alpha_{eff}=   C \alpha_S, 
$
which is what replaces $Q_c \alpha$ in Eq.~\eqref{QMpict}. From Fig.~\ref{chanpaton} we read $C=\sqrt{2}/3$ and find
\footnote{The factor $2$ arises from the two choices of the annihilating bilinear: given the symmetry of the tetraquark, we may call $c_1$ the annihilating $c$ quark and pair it to either $\bar c_1$ or $\bar c_2$; the spectroscopic factor is from~\eqref{fierz1}; in parenthesis $v\sigma(c\bar c\to q\bar q)$}:
\bea
&& \Gamma_5=\Gamma( {\cal T}\to M(c\bar q)+M(q\bar c))=
 2\cdot \frac{1}{6}\cdot \frac{2}{9}\cdot \left(\frac{4\pi\alpha_S^2}{3}  \frac{4}{m_{J/\psi}^2}\right)~|\Psi_{J/\psi}(0)|^2\cdot \xi \label{gammames}
\eea
Using Eq.~\eqref{fandpsi}, $\alpha_S=0.3$ and massless $q$,  we obtain
\be
 \Gamma_5= \frac{8\pi}{81}\left(\frac{\alpha_S}{f}\right)^2m_{J/\psi}\cdot\xi= 1.56~{\rm MeV} \cdot \xi
 \label{gammames2}
\ee
and 
\be
\Gamma({\cal T})= \Gamma_0+\Gamma_1 + 3\Gamma_5 =21\cdot \xi~{\rm MeV} 
\label{gamtot}
\ee
Eq.~\eqref{gammames2} gives the total decay rate into pseudoscalar and vector meson pairs. 
It is easy to see that the rate is shared between pseudoscalar and vector mesons in the ratio $1:3$~\cite{Becchi:2020mjz}. 

\emph{\bf Note.} We find $\Gamma_5\propto \alpha_S^2$. In Ref.~\cite{Anwar:2017toa} it was suggested that the leading tetraquark decay, at parton level, is ${\cal T}_{c_1c_2\bar c_3 \bar c_4} \to c_1\bar c_4 + g$, with the parton state evolving to the final hadrons with unit probability. This would give $\Gamma_5\propto \alpha_S$. We think however that the argument is not correct. To see this, we note that the parton state $c\bar c g$ corresponds to the cut along the line $a-a$ in Fig.~\ref{chanpaton}, with amplitude proportional to
\be
{\cal I}m\left(\frac{1}{q^2+i\epsilon}\right)=-\pi \delta(q^2)=0
\ee
since $q^2$ is strictly positive. The only non-vanishing parton cut is along $b-b$ and it corresponds to the decay ${\cal T}_{c_1c_2\bar c_3 \bar c_4} \to (c_1\bar c_4)_{{\bf 8}} + (q\bar q)_{{\bf 8}}$, which gives \eqref{gammames2}. Adding the contributions of light quark flavours ($u,~d,~s$) we obtain the total, inclusive rate into two open charm particles, with or without strangeness, $3\Gamma_5$.

Given the large $Q$-value available for the decay, the two, open charm particles could be accompanied by additional light mesons produced by the soft gluons radiated by the light quarks in the process of Fig.~\ref{chanpaton}. The use of $3\Gamma_5$ in \eqref{gamtot} is the same as approximating the full $\sigma(e^+ e^- \to {\rm hadrons})$ by $\sum_i\sigma(e^+ e^- \to q_i\bar q_i)$. 

A second possibility is that the annhilation of the spin $1$ color octet gives rise to two gluons, in place of the outgoing $q\bar q$ pair, parton decay ${\cal T}_{c_1c_2\bar c_3 \bar c_4} \to (c_1\bar c_4)_{{\bf 8}} + (g g)_{{\bf 8}}$. This is the same parton state considered in the annihilation of the spin $0$ color octet, item (4) of the list at the beginning of Sect.~\ref{calculation}, and it should be similarly suppressed.

\section{{\bf The value of $\xi$}}\label{value}

We estimate the ratio $\xi= |\Psi_{\cal T}(0)|^2/|\Psi_{J/\psi}(0)|^2$ by making use of numerical wave functions. These wave functions are obtained by means of a variational method with harmonic oscillator (h.o.) trial wave functions to solve the eigenvalue problem of a QCD Hamiltonian with One-Gluon-Exchange (OGE) interaction. 
This method was previously used in baryon and meson spectroscopy and tested on the reproduction of analytical and numerical (e.g. Ref. \cite{Barnes:2005pb}) results, both for the spectrum and the wave functions.

For the $J/\psi$ wave function the Hamiltonian we consider is that of the well-known relativized QM \cite{Godfrey:1985xj}, while in the tetraquark case the numerical wave function is extracted by means of the relativized diquark model of Refs.~\cite{Anwar:2017toa} and \cite{Bedolla:2019zwg}. We find
\footnote{The numerical wave functions of the $J/\psi$ and $0^{++}$ ground-state tetraquark can be fitted by simple harmonic oscillator wave functions, $\Psi({\bf r}) = \frac{1}{\pi^{3/4}} ~ \alpha_{\rm ho}^{3/2} \mbox{e}^{-\frac{1}{2}\alpha_{\rm ho}^2r^2}$ with $\alpha_{{\rm ho}; J/\psi} = 0.73$ GeV and $\alpha_{{\rm ho}; {\cal T}} = 1.3$ GeV. For ${\cal T}$, ${\bf r}$ is the distance of the c.o.m. of $cc$ and $\bar c\bar c$. The radial wave function in the origin is: $\Psi(0) = \frac{1}{\pi^{3/4}} ~ \alpha_{\rm ho}^{3/2}$.} 

  \be
 |\Psi_{(J/\psi,~{\rm h.o.})}(0)|^2=0.070~{\rm GeV}^3
 \ee
  \be
 |\Psi_{({\cal T},~{\rm h.o.})}(0)|^2=0.42~{\rm GeV}^3
 \ee 
The harmonic oscillator overlap probability for $J/\psi$ is smaller than the value obtained from $\Gamma(J/\psi \to \mu^+\mu^-)$, Eq.~\eqref{fandpsi2}.
 To estimate the ${\cal T}$ width, we take for $\xi$ the average of the two estimates and use their difference for the error
\bea
&&\xi_{{\rm h.o.}}=\frac{|\Psi_{({\cal T},~{\rm h.o.})}(0)|^2}{|\Psi_{(J/\psi,~{\rm h.o.})}(0)|^2}=6.0;~\xi_{{\rm h.o.},J/\psi}=\frac{|\Psi_{({\cal T},~{\rm h.o.})}(0)|^2}{|\Psi_{J/\psi}(0)|^2}=3.2\notag
\eea
\be
\xi=4.6\pm1.4\label{bestratio}
\ee

\section{ Tetraquark cross sections}

Combining Eqs.~\eqref{gam4mu} and \eqref{gamtot} we obtain, for $J^{PC}=0^{++}$:
\be
B_{4\mu}=B({\cal T}\to 4\mu)= 2.7~10^{-6}\label{b4mu}
\ee
and the cross section  upper bound
 \bea
 &&\sigma_{theo.}({\cal T}\to 4 \mu) 
  \leq \sigma(pp \to 2 J/\psi) B_{4\mu}=40~{\rm fb} \label{uplim}
 \eea
where $\sigma(pp \to 2 J/\psi)\simeq 15.2$ nb is the  two-$J/\psi$ production cross section measured by LHCb at 13 TeV \cite{Aaij:2016bqq}.


 \renewcommand{\arraystretch}{1.2}
 \begin{table}[htbp]
  \centering
 \begin{tabular}{|c|c|c|c|}
\hline
{\footnotesize $[cc][\bar c \bar c]$}& {\footnotesize Decay Channel }& $\begin{array}{c} BF ~{\rm in}\\ {\cal T} ~{\rm decay}\end{array}$ & {\footnotesize $\begin{array}{c} {\rm Cross ~section} \\ {\rm upper~limit ~(fb)}\end{array}$}   \\
\hline
 {\footnotesize $J=0^{++}$}
 &{\footnotesize ${\cal T} \to D^{(*)+} D^{(*)-}  \to e+\mu+\dots$} & {\footnotesize {$4.3 ~10^{-3}$}}  &  {\footnotesize $6.5\cdot 10^4$($65$~pb)}\\
  &{\footnotesize ${\cal T} \to D^{(*)0} {\bar D}^{(*)0}  \to e+\mu+\dots$} & {\footnotesize {$0.67~ 10^{-3}$}}  &  {\footnotesize $1.0\cdot 10^4$($10$~pb)}\\
   &{\footnotesize ${\cal T} \to 4\mu$} & {\footnotesize {$2.7~ 10^{-6}$}}  &  {\footnotesize $40$}\\
  \hline
   {\footnotesize $J=2^{++}$}&
{\footnotesize ${\cal T} \to D^{*+} {\bar D}^{*-}  \to e+\mu+\dots$}& {\footnotesize {$ 6.3~10^{-3}$}}  &  {\footnotesize {$ 9.6 \cdot 10^4  $($96$~pb)}}\\
  &{\footnotesize ${\cal T} \to D^{*0} {\bar D}^{*0}  \to e+\mu+\dots$} & {\footnotesize $ 0.98~10^{-3}$}  &  {\footnotesize $  1.5 \cdot 10^4 $($15$~pb)}\\
 &{\footnotesize ${\cal T} \to 4\mu$} & {\footnotesize {$1.6~ 10^{-5}$}}  &  {\footnotesize {$238$}}\\ 
\hline
\end{tabular}
\caption{\footnotesize{Upper limits  of two- and four-lepton cross sections via ${\cal T}$ production, estimated from the production cross sections of  $2~\psi(1S)$~(LHC, 13 TeV) \cite{Aaij:2016bqq}.}
}
\label{tab:width}
\end{table}

We focus on the $e\mu$ inclusive channel and give
in Tab.  \ref{tab:width} the upper limits to $\sigma_{theo.}({\cal T} \to 2 D^{(*)}_q\to  e~\mu+\dots)$, calculated as
   \bea
&& \sigma_{theo.}({\cal T} \to 2 D^{(*)}_q\to e \mu+\dots) = \sigma(pp \to {\cal T+\dots})  BF({\cal T} \to 2 D^{(*)}_q\to e \mu+\dots)
 \nonumber\\
&& \leq \sigma(pp \to 2 J/\psi+\dots) BF({\cal T} \to 2 D^{(*)}_q\to e \mu+\dots)
\label{Spred}
 \eea 

The largest part of the signal (the total signal for $J^{PC}=2^{++}$) arises from the decay of  ${\cal T}$ into a pair of vector mesons. Vector particles decay promptly into a pseudoscalar plus a soft pion or photon(s) and contribute to the signal on the same basis as the pseudoscalars.

 In conclusion,  production in the $4\mu$ channel and decay rates that we estimate for the $cc\bar c\bar c$ tetraquarks are tantalizingly similar to the preliminary results presented by the LHCb Collaboration~\cite{Aaij:2020fnh}. The meson-meson channel with the $e\mu$ signature may provide an additional, complementary tool to identify and study the spectacular, exotic $cc\bar c\bar c$ tetraquarks. 

We thank Sheldon Stone, for an enlightening discussion and advice on a preliminary, March 2020, version of these notes, Michelangelo Mangano for an interesting exchange on the hadronization of the light quark pair and Liupan An for interesting correspondence on her seminar at CERN.

\end{document}